\documentclass[[journal=jacsat,manuscript=article]{achemso}
\usepackage[version=3]{mhchem}
\usepackage{color}
\usepackage{siunitx}

\title{Vacancy order and physical properties of a ternary compound \ce{Fe_{0.68}Pd_{0.80}Te} with $\alpha$-Fe$_{1+x}$Te-type structure}

\author{Bingxian Shi,$^{1,3,\dag}$ Manyu Wang,$^{1,\dag}$ Chenglin Shang,$^{1,3,\dag}$ Yanyan Geng,$^{1}$ Yangyang Fan,$^{1,3}$ Xiaoxiao Pei,$^{2}$ Li Huang,$^{2}$ Daye Xu,$^{1,3}$ Juanjuan Liu,$^{1,3}$ Jinchen Wang,$^{1,3}$ Hongxia Zhang,$^{1,3}$ Hongliang Wang,$^{4}$ Lijie Hao,$^{4}$ Rui Xu,$^{1}$ Zhongyi Lu,$^{1,3}$ Zhihai Cheng,$^{1*}$ and Peng Cheng$^{1,3*}$}

\email{zhihaicheng@ruc.edu.cn, pcheng@ruc.edu.cn}
	
\affiliation[RUC]
{$^{1}$ Key Laboratory of Quantum State Construction and Manipulation (Ministry of Education),
    School of Physics, Renmin University of China, Beijing 100872, China\\
	$^{2}$Beijing National Laboratory for Condensed Matter Physics, Institute of Physics, Chinese Academy of Sciences, Beijing 100190, China\\
	$^{3}$Laboratory for Neutron Scattering, School of Physics, Renmin University of China, Beijing 100872, China\\
	$^{4}$China Institute of Atomic Energy, PO Box-275-30, Beijing 102413, China
}

\begin{document}

\begin{abstract}
We report the identification and characterization of a new compound \ce{Fe_{0.68}Pd_{0.80}Te} with $\alpha$-Fe$_{1+x}$Te prototype structure. Different from the Fe-square net and minor occupancy of interstitial Fe-sites in Fe$_{1+x}$Te, \ce{Fe_{0.68}Pd_{0.80}Te} is featured by Pd-square net and near 68\% occupancy of the corresponding interstitial Fe-sites. Furthermore, non-contact atomic force microscopy and X-ray diffraction provide evidences for the existence of a 3$\times$3$\times$3 Pd-vacancy order in this layered material. A spin-glass ground state below $T_g$$\sim$40~K is identified via magnetic characterization. 
Electrical transport measurements show that \ce{Fe_{0.68}Pd_{0.80}Te} is a semiconductor with very small band gap below 10~meV. It has weak negative magnetoresistance and hole-like charge carriers below room temperature. Our results demonstrate its potentials for further exploring various quantum phenomena.
\end{abstract}

\section{Introduction}

Iron chalcogenides have draw significant attentions for the discoveries of many unconventional superconductors in them. The famous examples includes $\alpha$-FeSe\cite{FeSe,Cava}, FeTe$_{1-x}$Se$_x$\cite{FeTeSe}, FeS\cite{FeS} and K$_2$Fe$_4$Se$_5$\cite{KFeSe}. They are widely known as Fe-based superconductors, which is the second high-temperature superconducting family after cuprate whose superconducting mechanism is believed to be unconventional and closely related to magnetism\cite{PCDai,Review2022}. The first three all have tetragonal PbO-type phase, same as $\alpha$-Fe$_{1+x}$Te. Fe$_{1+x}$Te is initially accepted as an antiferromagnet with T$_N$$\sim$65~K\cite{FeTeorder,Lishiliang} and can only become superconducting either by Se-doping or in the form
of thin-film under tensile stress\cite{FeTeSe,FeTefilm}. Interestingly, a very recent discovery reveals that if one can remove all the interstitial Fe atoms, then stoichiometric FeTe is actually a superconductor with a critical temperature of 13.5~K\cite{FeTe2026}. Furthermore, FeTe$_{1-x}$Se$_x$ (x=0.45) is also identified to host Dirac-cone–type surface states at the Fermi level and may be a topological superconductor with exotic Majorana states\cite{Zhang2018,Gao,Machida2019,Gao2}. Additionally, FeTe has two-dimensional (2D) van der Waals layered structure. Through synthesizing heterostructures formed by stacking ferromagnetic topological insulator Cr-doped (Bi,Sb)$_2$Te$_3$ and FeTe, interface-induced chiral topological superconductivity was observed recently\cite{Science2024}. The above discoveries make Fe$_{1+x}$Te receive longstanding research interests.

On the other hand, atomic vacancies as the simplest form of point defects widely exist in various materials, but only few of them may have vacancy-ordering phase. The K$_2$Fe$_4$Se$_5$ superconductor mentioned above has an insulating $\sqrt{5}\times\sqrt{5}$ iron vacancy-ordering phase that coexists with the superconducting phase at the nanoscale\cite{ZAVA,Bao2011,LJQ,Bao245review,245RMP}. Although its vacancy order may not be responsible for superconductivity, there was a report about superconductivity emerging by suppressing vacancy ordering in Ir$_{16}$Sb$_{18}$ which is named as "correlated vacancies"\cite{Qi}. Besides, vacancy order is also proposed to be related to many other intriguing magnetic and transport properties\cite{V1,V2,V3}.

In this work, we report the discovery of a ternary compound \ce{Fe_{0.68}Pd_{0.80}Te}. Its crystal structure resembles that of tetragonal Fe$_{1+x}$Te. A short-range 3$\times$3$\times$3 Pd-vacancy order is observed from both single crystal X-ray diffraction and non-contact atomic force microscopy. Magnetization measurements reveal a spin-glass ground state below $T_g$=40~K. Transport data shows that \ce{Fe_{0.68}Pd_{0.80}Te} is a semiconductor with very small band gap and hole-like charge carriers. Our findings establish \ce{Fe_{1-x}Pd_{1-y}Te} as a system for further exploring emergent quantum phenomena.

\section{Experimental Section}

Single crystals of \ce{Fe_{0.68}Pd_{0.80}Te} were grown by the high-temperature solution growth method similar as growing \ce{FePd2Te2}, a 2D ferromagnetic metal that we have recently discovered\cite{FePdTe}. The crucial point in obtaining crystals with different Fe-Pd-Te phases is the initial elements molar ratio. For \ce{Fe_{0.68}Pd_{0.80}Te}, Pd, Fe and Te powders were mixed with a molar ratio of 1.5:2:2 and placed in an alumina crucible, then sealed in an evacuated quartz tube. The tube was heated up to \SI{800}{\celsius} then maintained at this temperature for two days. Then it was slowly cooled to \SI{600}{\celsius} at a rate of \SI{2}{\celsius}/h and keep at this temperature for another day before furnace-cooled to room temperature. Plate-like large crystals with typical plane-size of 5~mm $\times$ 5~mm could be obtained. These crystals are stable in air. It is found that the fabrication of phase-pure polycrystalline samples of \ce{Fe_{0.68}Pd_{0.80}Te} by high-temperature annealing of the mixture of elements is very difficult. Notable impurity phases such as Fe$_{1+x}$Te, PdTe, PdTe$_2$ or FePd$_2$Te$_2$ would appear in the product.

X-ray diffraction (XRD) data of the single crystals were collected at room temperature from a Bruker D8 VENTURE PHOTO II diffractometer equipped with multilayer mirror monochromatized Mo K$_{\alpha}$ (${\lambda}$ = 0.71073 {\AA}) radiation and Bruker D8 Advance X-ray diffractometer using Cu K$_{\alpha}$ radiation. The structural solution and precession photos for \ce{Fe_{0.68}Pd_{0.80}Te} were obtained using the APEX3 program. The final refinement was completed with the SHELXL suite of programs. The elemental composition of single crystals were examined with energy dispersive x-ray spectroscopy (EDS, Oxford X-Max 50). Magnetization data were collected on a Quantum Design Magnetic Property Measurement System (MPMS). The electrical transport measurements were carried out on a Quantum Design Physical Property Measurements System (PPMS-14T) with four-probe method. The single crystal neutron diffraction experiment was carried out on Xingzhi cold neutron triple-axis spectrometer at the China Advanced Research Reactor (CARR)\cite{XingZhi}. The incident neutron energy was fixed at 16~meV.

The non-contact atomic force microscopy (nc-AFM) measurements were performed in a commercial LT-STM system equipped with an STM/qPlus sensor at 4.5~K. \ce{Fe_{0.68}Pd_{0.80}Te} single crystals were cleaved at liquid nitrogen temperature in ultrahigh vacuum. The nc-AFM images were recorded by measuring the frequency shift of the qPlus resonator (sensor frequency $f_0$=30~kHz, Q=53000) in constant-height mode with an oscillation amplitude of 200~pm.

For scanning tunneling microscopy (STM) experiment, \ce{Fe_{0.68}Pd_{0.80}Te} single crystals were cleaved at room temperature in ultrahigh vacuum at a base pressure of 2$\times$10$^{-10}$~Torr, then directly transferred to the cryogen-free variable-temperature STM system (PanScan Freedom, RHK). Chemically etched Pt-Ir tips were used for STM measurement in constant current mode. The tips were calibrated on a clean Ag(111) surface. Gwyddion was used for STM data analysis.

\section{Results and Discussion}

\subsection{Crystal structure}

\begin{figure}
	\centerline{\includegraphics[width=\textwidth]{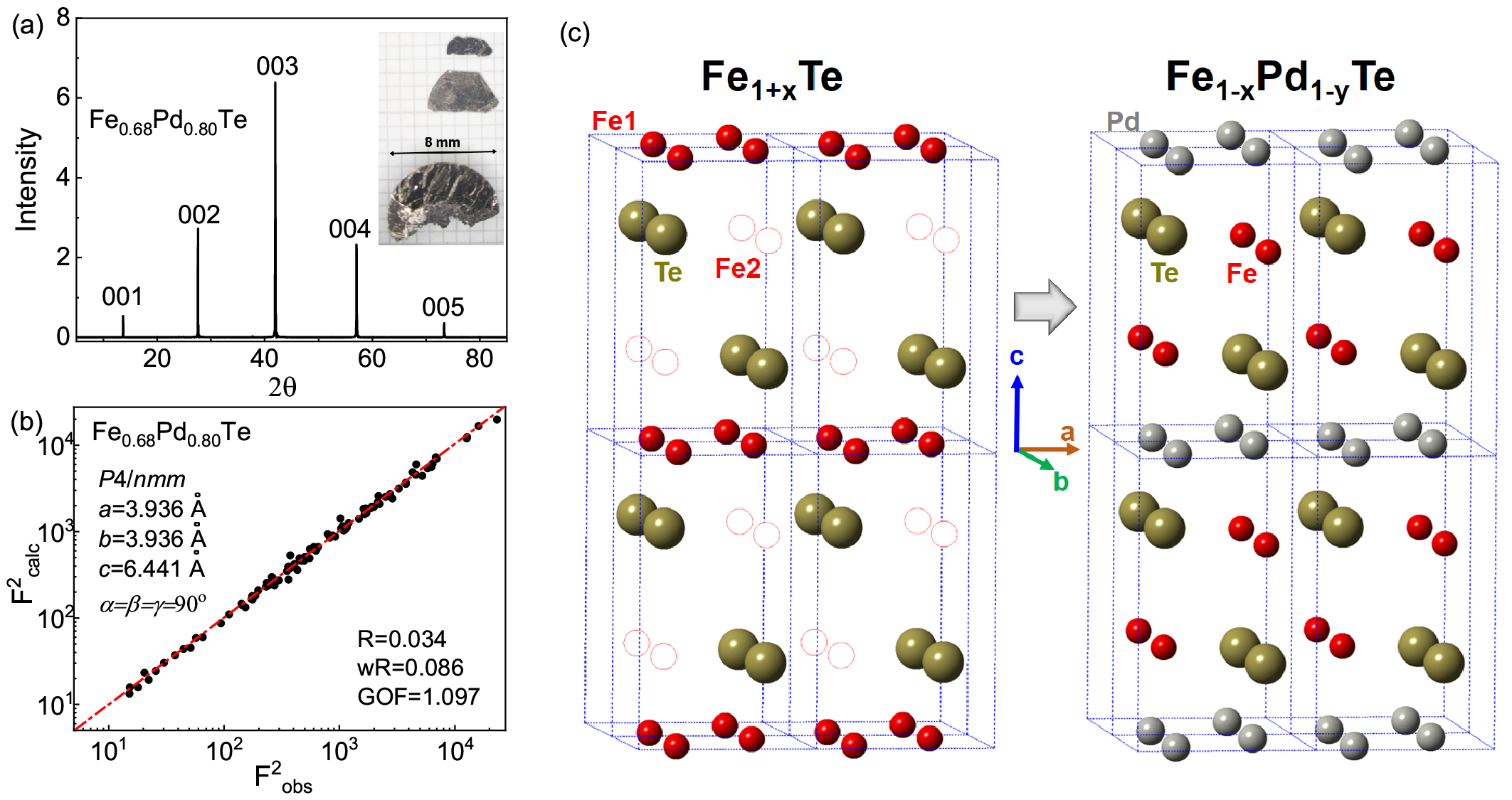}} \vspace*{-0.3cm}
	\caption{Crystal structure of \ce{Fe_{0.68}Pd_{0.80}Te}. (a) X-ray diffraction patterns from the $ab$-plane of \ce{Fe_{0.68}Pd_{0.80}Te}. The inset shows the photos of as-grown single crystals. (b) Refinement result of lattice Bragg peaks obtained from single crystal X-ray diffraction. (c) Crystal structure of Fe$_{1+x}$Te and \ce{Fe_{1-x}Pd_{1-y}Te}. The $x$ value in the former denotes the occupancy of interstitial site Fe2, while the $x$ and $y$ values in the latter denote the Fe- and Pd-site vacancy respectively.} 
\end{figure}

The discovery of Fe$_{0.68}$Pd$_{0.80}$Te emerged from our search for 2D magnetic materials in Fe-Pd-Te ternary compounds\cite{FePdTe}. After the growth of large-size plate-like single crystals shown in the inset of Fig. 1(a), their chemical formula is found to be Fe$_{0.68}$Pd$_{0.80}$Te which is determined by the EDS analysis on the chemical compositions. The standard relative uncertainty for the occupancies of Fe and Pd is estimated to be 6\%. Then single crystal X-ray diffraction experiments and structural analysis/refinement were performed as described in the previous section. The result is shown in Fig. 1(b), Table S1 and S2 of the supporting information. It should be mentioned that the chemical formula refined from single crystal XRD is slightly different from that determined by EDS, which should be due to the incomplete description of the superlattice peaks induced by vacancy orders in XRD refinement\cite{FePdxTe2}. The chemical formula Fe$_{0.68}$Pd$_{0.80}$Te determined by EDS should be accurate as it is consistent with the nc-AFM data and makes the magnetization data analysis physically reasonable, which will be introduced later.

Fe$_{0.68}$Pd$_{0.80}$Te adopts a tetragonal structure with $P4/nmm$ space group (No. 129) same as Fe$_{1+x}$Te. We will first briefly describe the structure of Fe$_{1+x}$Te. As shown in Fig. 1(c), the iron atoms in Fe$_{1+x}$Te have two different Wyckoff positions. Fe1 forms a square net with full occupancy and a stack of edge-sharing FeTe$_4$-tetrahedra layers combined with adjacent Te-atoms. This layer is believed to be crucial for emerging unconventional superconductivity. In addition, numerous previous investigations have shown that stoichiometric Fe$_{1+x}$Te (x=0) is hard to obtain and usually there are excess Fe partially occupies the interstitial site Fe2\cite{FeTeorder,Lishiliang,Inter1,Inter2,Inter3,Inter4}. However, the occupancy of Fe2 site is usually very low with the value of x no more than 0.15. 

From Fe$_{1+x}$Te to Fe$_{0.68}$Pd$_{0.80}$Te, the major differences are that Pd-atoms occupy the Fe1 site with occupancy of $\sim$80\% while all the Fe-atoms migrate to the interstitial site Fe2 with occupancy of $\sim$68\%. Therefore, although Fe$_{0.68}$Pd$_{0.80}$Te has the same structural framework as that of Fe$_{1+x}$Te, the significant change of atomic position and occupancy mean that Fe$_{0.68}$Pd$_{0.80}$Te cannot be simply viewed as a Pd-doped FeTe system. Owing to the large difference in the X-ray scattering length between Fe and Pd, the single-crystal XRD data has enough resolution to distinguish the proposed structure from other models with alternative atomic arrangements. Furthermore, the powder XRD data could also be well refined by the proposed crystal model as shown in Fig. S1. For the powder XRD data, the intensities of (00L) Bragg peaks are much stronger than the calculated intensities and the preferred crystal orientation along (00L) must be used in the Rietveld refinement. This suggests that Fe$_{0.68}$Pd$_{0.80}$Te is a two-dimensional crystal system. Additionally, both the $a$- and $c$-lattice constants of Fe$_{0.68}$Pd$_{0.80}$Te expand by approximately 3\% relative to Fe$_{1+x}$Te. This is naturally explained by the introduction of Pd atoms with larger ion radius. It should be noted that there is a tetragonal-to-monoclinic structural phase transition in Fe$_{1+x}$Te\cite{FeTeorder,Lishiliang}, while this transition does not exist in Fe$_{0.68}$Pd$_{0.80}$Te as no clues are detected from both resistivity data and low-temperature neutron diffraction. 

The tetragonal crystal structure for Fe$_{1+x}$Te cannot be stabilized with large occupancy of interstitial Fe. However for Fe$_{0.68}$Pd$_{0.80}$Te, since Pd has different electronegativity and ion radius from that of Fe, the introduction of Pd will significantly alter the chemical bonding, local coordination environment and band structures in this system, which somehow make the 68\% occupancy of interstitial Fe possible. Future theoretical calculations could be stimulated for further investigating the underlying mechanism.

\subsection{Vacancy order}
Fig. 2(a-c) show the precession photos extracted from single-crystal XRD data of Fe$_{0.68}$Pd$_{0.80}$Te. For the ($H$0$L$) reciprocal plane, besides the lattice Bragg reflections at ($H$, 0, $L$) ($H$, $L$=$\pm$1,$\pm$2...), additional superlattice Bragg peaks appear at ($H$$\pm$1/3, 0, $L$$\pm$1/3) highlighted by the rectangular marker. Similarly, superlattice Bragg peaks also appear at (0, $K$$\pm$1/3, $L$$\pm$1/3). No superlattice peaks can be seen on the ($H$$K$0) plane and one would notice the structural extinction effect on the ($H$, $K$, 0) ($H$, $K$=odd numbers) peaks. The above observations suggest a 3$\times$3$\times$3 superstructure exists in Fe$_{0.68}$Pd$_{0.80}$Te. Since there are substantial Fe- and Pd-vacancies in this material, this superstructure is most likely caused by the atomic vacancy orders.

\begin{figure}[tbp]
	\centerline{\includegraphics[width=\textwidth]{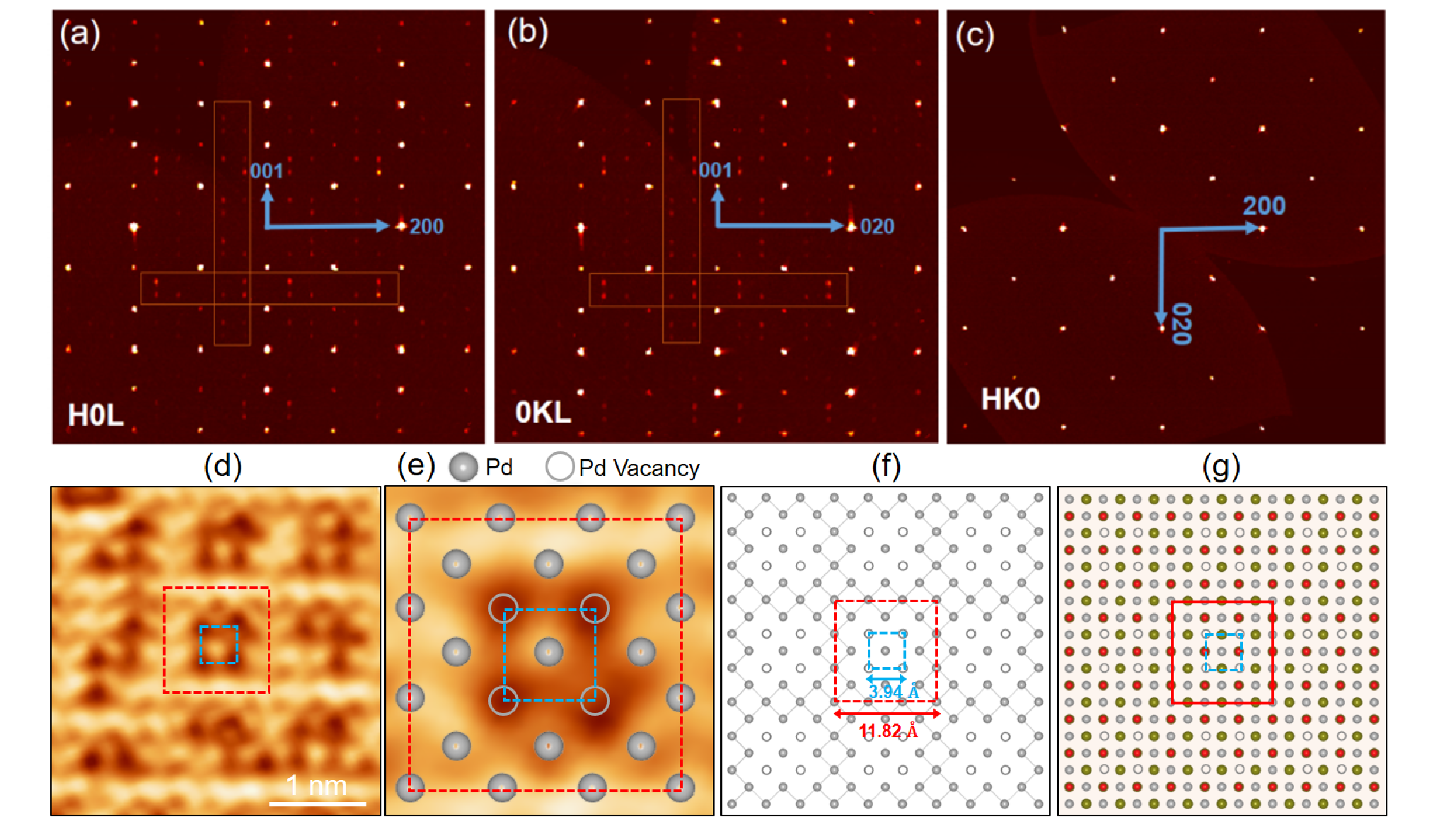}} \vspace*{-0.3cm}
	\caption{Vacancy order in \ce{Fe_{0.68}Pd_{0.80}Te}. (a-c) Precession photos generated from single crystal X-ray diffraction data which show the reciprocal lattice views on ($H$0$L$), (0$KL$) and ($HK$0) respectively. (d) Atomic-resolved nc-AFM image on the cleavage plane of \ce{Fe_{0.68}Pd_{0.80}Te}. (e) Zoom-in nc-AFM image and the schematic model of the 3$\times$3 Pd-vacancy order. Schematic models of the vacancy order in Pd-middle-sublayer and FeTe-Pd-FeTe trilayer are shown in (f) and (g) respectively. The extended 2D unit cell is illustrated using red square. The blue square which connects the four nearest Pd-vacancies also represent the size of initial 2D unit cell in the $ab$-plane.}
\end{figure}

As a local probe with atomic resolution, non-contact atomic force microscopy (nc-AFM) could directly observe the distribution of atoms or vacancies on the 2D surface. Fig. 2(d,e) shows the nc-AFM images on the cleaved surface of Fe$_{0.68}$Pd$_{0.80}$Te single crystal. Assuming the bright regions represent the occupancy of Pd atoms and the dark regions represent the Pd vacancies, then a 2D 3$\times$3 ordered pattern of Pd-vacancies as illustrated in Fig. 2(e,f) coincides well with the atomically resolved nc‑AFM images. Not only the atomic distances shown in Fig. 2(f) that determined by XRD agrees well with that in nc-AFM image, the zigzag patterns generated by connecting the nearest Pd-Pd bondings in Fig. 2(f) also clearly present in the nc-AFM image [Fig. 2(d)]. More nc-AFM and STM images are shown in Fig. S2 which are consistent with the above conclusion. We should mention that since the Fe-square net in the 2D $ab$-plane has different atomic distances [Fig. 2(g)], this pattern cannot be explained by the Fe-vacancy order. 

Therefore, nc-AFM reveals the 3$\times$3 Pd-vacancy order pattern in the 2D Pd-square net. According to this vacancy distribution, the occupancy on Pd-site should be 7/9 (~78\%), which agrees well with 80\% determined from EDS and provides further verification on our conclusion. This 2D vacancy pattern should also have modulations along the $c$-axis, which is an ABC-stacked trilayers style as in 2D materials. Fig. S3 presents a schematic illustration of the three-dimensional 3$\times$3$\times$3 supercell of \ce{Fe_{0.68}Pd_{0.80}Te} based on the proposed Pd-vacancy order. Then all the superlattice Bragg peaks observed in the precession photos can be well explained and reach consistency with the Pd-vacancy order shown in Fig. 2.

Next, it should be mentioned that the superlattice Bragg peaks are rather weak and seem to be notable only in the low-$Q$ area, as seen from the zoom-out views of precession photos shown in Fig. S4 in the supporting information. This resembles that of FeGe, which is a kagome metal with superlattice Bragg peaks originated from short-range charge-density-wave (CDW) order\cite{FeGe,FeGeChongqing}. In FeGe, when tuning the CDW order to a long-range one by post-annealing, the superlattice Bragg peaks would be clearly visible in the whole $Q$-range\cite{FeGeChongqing}. Therefore, we speculate that the vacancy order in Fe$_{0.68}$Pd$_{0.80}$Te may also be short-range. The weak superlattice Bragg peaks make the complete refinement of single-crystal XRD data including the supercell difficult and the XRD refinement result presented in this work does not include the description of the superlattice structure. However our current experimental result has provided strong evidence for the existence of 3$\times$3$\times$3 Pd-vacancy order in Fe$_{0.68}$Pd$_{0.80}$Te. Recently, superlattice Bragg peaks were also identified in two other Fe-Pd-Te ternary compounds FePd$_2$Te$_2$ and FePd$_{2.5}$Te$_2$\cite{FePdxTe2}. The details of their supercells were not known but proposed to be associated with the possible complex ordering of Pd/Fe atoms or the interstitial Pd in the van der Waals gaps\cite{FePdxTe2}. Our result suggests that vacancy order could be another possible explantion to understand the mysterious superlattice Bragg peaks in Fe-Pd-Te material family.

Different from the vacancy order in K$_2$Fe$_4$Se$_5$ that only has $\sqrt{5}\times\sqrt{5}$ modulation along $ab$-plane\cite{Bao2011,Bao245review}, the vacancy order in \ce{Fe_{0.68}Pd_{0.80}Te} also has modulation along the $c$-axis, possibly due to the enhanced interlayer coupling. One may notice that the Fe-deficiency determined by EDS is close to 1/3, it is interesting to speculate that Fe-vacancy order may also possibly exist which requires further investigation with advanced probes. On the other hand, the relationship between vacancy order, band structures and physical properties in \ce{Fe_{0.68}Pd_{0.80}Te} is another interesting topic for further theoretical investigations.

\subsection{Magnetic and transport properties}

\begin{figure}[tbp]
	\centerline{\includegraphics[width=\textwidth]{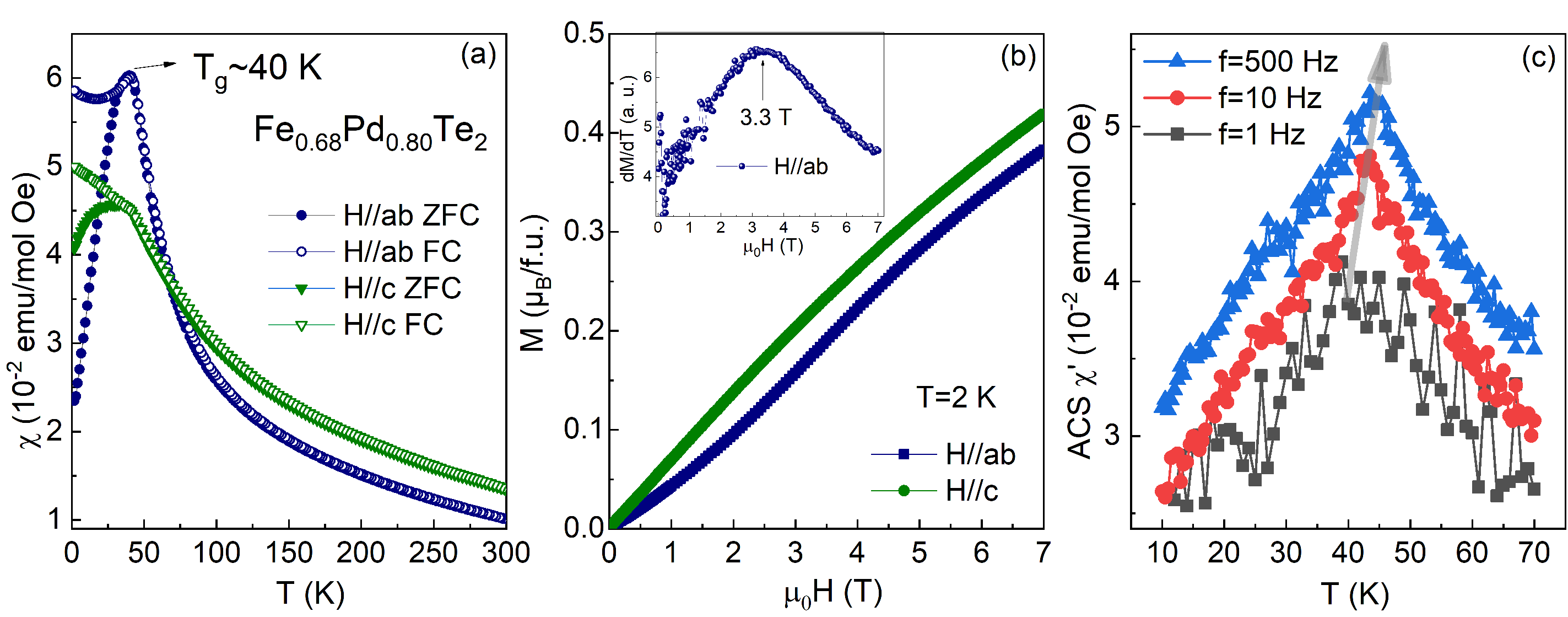}} \vspace*{-0.3cm}
	\caption {Magnetic properties of \ce{Fe_{0.68}Pd_{0.80}Te}. (a) Temperature-dependent magnetic susceptibilities under magnetic field of 0.1~T applied along H$\parallel$$ab$ or H$\parallel$$c$. (b) Isothermal magnetization at T=2~K. The inset shows the dM/dT curve for data along H$\parallel$$ab$. (c) The temperature-dependent AC susceptibilities measured under an oscillated	AC field of 5.0 Oe applied along H$\parallel$$ab$. The peak in real part $\chi$' shows a shift with increasing frequencies.}
\end{figure}

Magnetization and resistivity measurements were performed to reveal the physical properties of \ce{Fe_{0.68}Pd_{0.80}Te}. Fig. 3(a) presents the temperature-dependent magnetic susceptibility data $\chi$(T). For both H$\parallel$$ab$ and H$\parallel$$c$, zero-field-cooling (ZFC) $\chi$(T) data gradually drops below 40~K, which suggests an antiferromagnetic-like transition. Nevertheless, the field-cooling (FC) $\chi$(T) data only show a weak kink across 40~K and there is a large bifurcation between ZFC and FC $\chi$(T) curves below 40~K. This feature strongly indicates a spin-glass state below $T_g$$\sim$40~K\cite{Li_2018,Pan_2022,Huangjiale}. Furthermore, the Curie-Weiss (CW) fit on the $\chi$(T) data in the paramagnetic state (250~K-300~K) yields $\theta^{ab}_{CW}$=32~K and $\mu^{ab}_{eff}$=4.53~$\mu_{B}$/Fe for H$\parallel$ab, $\theta^{c}_{CW}$=37~K and $\mu^{c}_{eff}$=4.49~$\mu_{B}$/Fe for H$\parallel$$c$. The derived effective moment values are close to an $S$=2 high-spin state for Fe$^{2+}$. The positive $\theta_{CW}$ values suggest dominate ferromagnetic interactions and there might be a competition between ferromagnetic/antiferromagnetic interactions which leads to the spin-glass ground state.

\begin{figure}[tbp]
	\centerline{\includegraphics[width=\textwidth]{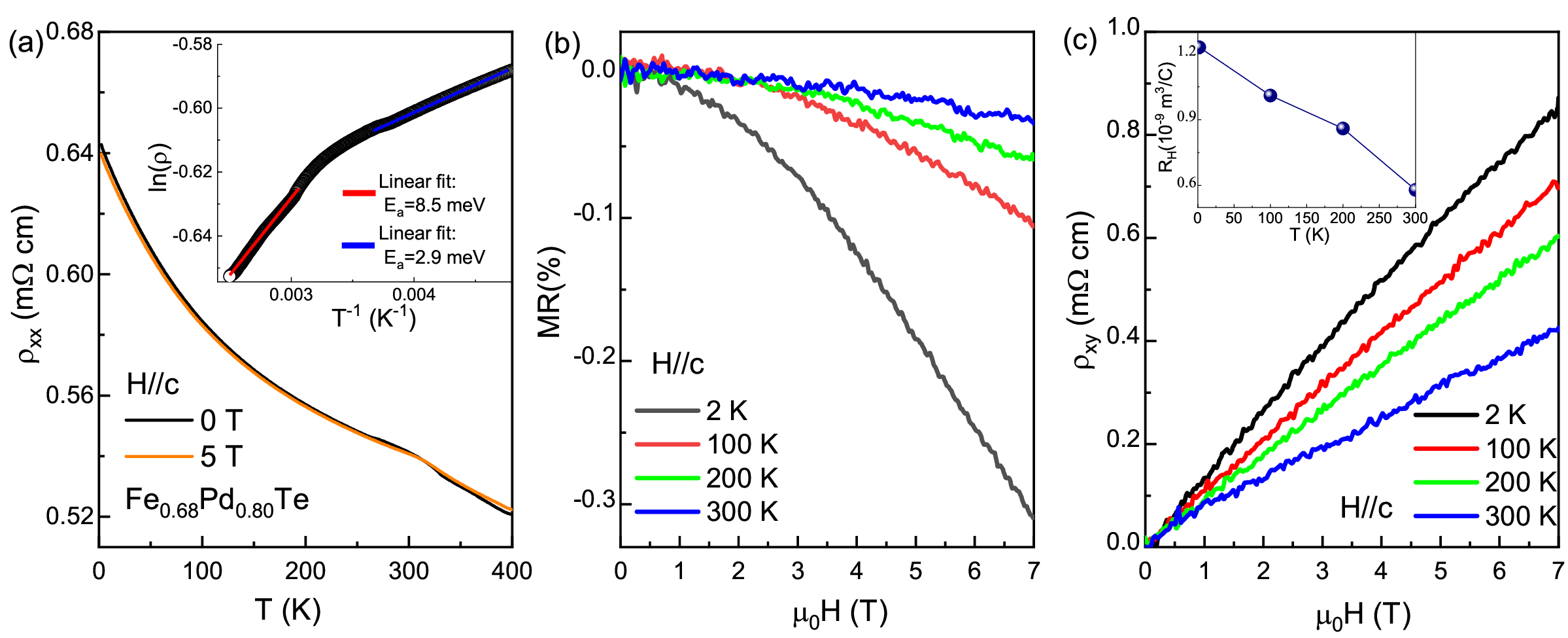}} \vspace*{-0.3cm}
	\caption {Electrical transport properties of \ce{Fe_{0.68}Pd_{0.80}Te}. (a) The temperature-depdendent resistivity under magnetic field of 0~T and 5~T. The inset shows a plot of ln($\rho$) as a function of T$^{-1}$. (b) Magnetoresistance at selected temperatures. (c) The field-dependent Hall resistivity $\rho_{xy}$ at selected temperatures. The inset shows the temperature-dependent Hall coefficients obtained from linear fit of $\rho_{xy}$(H).} 
\end{figure}

To further confirm the spin-glass transition, AC susceptibilities measured under different AC frequencies ($f$) are shown in Fig. 3(c). With increasing $f$ from 1~Hz to 500~Hz, the peak position of the real component $\chi$' shifts systematically toward higher temperatures by several kelvins, consistent with the spin frozen state below spin-glass transition at $T_g$$\sim$40~K\cite{Pan_2022}. Neutron diffraction measurements were also performed on a single crystal of \ce{Fe_{0.68}Pd_{0.80}Te} to search for possible magnetic orders below $T_g$. After scanning all possible Bragg peaks in the (H0L) plane, only the intensity of [001] Bragg peak shows certain increase at 3.5~K comparing with that at 65~K but no well-defined order parameter behavior is observed (Fig. S5). This observation confirms the absence of uniform long-range magnetic order and points toward short-range cluster spin-glass correlations throughout the material.

Our results also suggest that \ce{Fe_{0.68}Pd_{0.80}Te} has an in-plane magnetic anisotropy. Firstly, the susceptibility drop below $T_g$ is far sharper for H$\parallel$$ab$ than for H$\parallel$$c$, demonstrating the spins preferentially lie within the $ab$ basal plane. Secondly, as shown in the $M(H)$ and $dM/dT$ curves in Fig. 3(b), a metamagnetic transition at 3.3~T is identified only under H$\parallel$$ab$. This is consistent with the fact that metamagnetic transitions usually exist when the field is applied along the easy-axis/easy-plane. Thirdly, in neutron diffraction data, magnetic contributions on the Bragg peaks exist only in [001] but not on any [H00] peaks, which also implies the in-plane magnetic anisotropy according to the magnetic neutron scattering rule.

Fig. 4(a) displays the temperature-dependent longitudinal resistivity $\rho_{xx}$(T). $\rho_{xx}$(T) increases with decreasing temperature which is a semiconducting behavior. The inset of Fig. 4(a) plots ln($\rho$) versus $T^{-1}$, revealing two distinct temperature windows (200–250~K and 330–400~K) with quasi-linear behavior, which can be fitted by the thermal activation model defined by $\rho$=$\rho_0$exp($E_a$/2$k_B$T). $E_a$ is the energy gap and $k_B$ is the Boltzmann constant. The fitting results shown by the red and blue solid lines give $E_a$=8.5~meV and 2.9~meV respectively. Therefore \ce{Fe_{0.68}Pd_{0.80}Te} is likely a semiconductor with very small band gap. On the other hand, one may notice that the absolute value of resistivity is only around 0.6~m$\Omega$~cm, which is comparable to that of Fe$_{1+x}$Te\cite{FeTeSe} and even smaller than many parent compounds of Fe-based superconductors such as LaFeAsO\cite{Hosono} and BaFe$_2$As$_2$\cite{BaFe2As2}. These Fe-based parent compounds are well-known bad metals with similar weak semiconducting behavior. Therefore \ce{Fe_{0.68}Pd_{0.80}Te} is also possibly a bad metal as Fe$_{1+x}$Te and other Fe-based compounds\cite{FeTeSe,SUN,FeGeSb}. 

The $\rho_{xx}$(T) curve does not exhibit any anomaly at $T_g$=40~K, which might be due to the failure to develop long-range magnetic order. Interestingly, there is an anomaly at around 320~K. This anomaly might originate from a vacancy order to disorder transition. Possibly the atomic vacancy order would disappear at higher temperatures, as similar order to disorder transitions were observed in K$_2$Fe$_4$Se$_5$\cite{Bao2011} and other Fe-Pd-Te compounds\cite{FePdxTe2}. Below 320~K, the magnetoresistance (MR) become negative down to 2~K. This weak negative MR can be observed from both Fig. 4(a) and (b). For one thing, the negative MR may result from the reduced spin scattering by applying magnetic field, as it gets much stronger below 40~K. For another, it might also be related to the disorder or localization effect\cite{MR}.

Hall effect and transverse resistivity $\rho_{xy}$ data for \ce{Fe_{0.68}Pd_{0.80}Te} are shown in Fig. 4(c). $\rho_{xy}$ exhibits quasi-linear field dependence within the instrumental resolution. Their slopes are positive for all measured temperatures indicating hole-like charge carriers in this compound. Through linear fits of $\rho_{xy}$(H), the Hall coefficients $R_H$ at different temperatures were obtained and presented in the inset of Fig. 4(c). $R_H$ decreases quickly with increasing temperature, which agrees well with the behavior in semiconductors.

Prior literatures on Fe–Pd–Te ternary inorganic phases are very rare, with only one report investigating Pd-doped FeTe, documenting the evolution from tetragonal FeTe-phase to hexagonal PdTe-phase\cite{PNAS}. In the inorganic crystal structure database (ICSD), there had been no record in any Fe-Pd-Te ternary phase until we have recently discovered a 2D ferromagnetic metal \ce{FePd2Te2} with Curie temperature of 183~K\cite{FePdTe}. Then 2D ferromagnetic metals FePd$_{2.3}$Te$_2$ and Fe$_{0.9}$Pd$_{2.5}$Te$_2$ were also reported with Curie temperatures of 112~K and 98~K respectively\cite{FePdxTe2}. In this work, by tuning the relative Fe-Pd-Te compositions, a new chemical phase \ce{Fe_{0.68}Pd_{0.80}Te} is identified. Although the crystal structures of all these Fe-Pd-Te compounds can be viewed as derivatives of the tetragonal Fe$_{1+x}$Te prototype, Fe$_x$Pd$_y$Te$_2$ (x$\sim$1, y$\sim$2) is featured by one-dimensional Fe or Pd zigzag chains while \ce{Fe_{0.68}Pd_{0.80}Te} is more closely related to Fe$_{1+x}$Te with square net of transitional metal atoms. Besides, \ce{Fe_{0.68}Pd_{0.80}Te} possesses a spin-glass magnetic state which makes it quite unique in all the Fe-Pd-Te ternary compounds mentioned above.

Finally, searching for new emergent phenomena such as superconductivity may also be stimulated in chemical-doped or pressurized \ce{Fe_{0.68}Pd_{0.80}Te}, because of its structural connection with the parent compound of Fe-based superconductor Fe$_{1+x}$Te. Recently, superconductivity at 30.4~K is realized in Fe$_{1.1}$Se with 11\% interstitial Fe ions\cite{FeSe2026}. Indeed, we have observed filamentary superconductivity with superconducting critical temperature of 7~K in some \ce{Fe_{1-x}Pd_{1-y}Te} samples. This $T_C$ is higher than PdTe\cite{PdTe} or PdTe$_2$\cite{PdTe2} superconductors which might constitute the impurities. This observation still need further confirmation and beyond the scope of this work. On the other hand, in recent years, there have been many efforts in synthesizing heterostructures using FeTe or FeTe$_{1-x}$Se$_x$ with other Te-based compounds, aiming to explore proximity induced topological superconductivity or other quantum effects\cite{Science2024,He2014,MnTe,WHH,HeteroPRM,HeteroFGT}. Since the lattice of \ce{Fe_{0.68}Pd_{0.80}Te} also has a 2D layered structure and matches that of FeTe, it is an ideal member for such application.

\section{Conclusion}

In summary, a ternary compound \ce{Fe_{0.68}Pd_{0.80}Te} with $\alpha$-Fe$_{1+x}$Te-type structure is reported. Comparing with Fe$_{1+x}$Te, \ce{Fe_{0.68}Pd_{0.80}Te} is characterized by Pd-square net and 68\% occupancy of Fe atoms on the interstitial Fe-sites. The existence of a 3$\times$3$\times$3 Pd-vacancy order is evidenced by nc-AFM and XRD. \ce{Fe_{0.68}Pd_{0.80}Te} has a spin-glass magnetic ground state below $T_g$$\sim$40~K. Transport measurements suggest it is a semiconductor with small band gap of below 10~meV and hole-like charge carriers. \ce{Fe_{0.68}Pd_{0.80}Te} may serve as a material platform for further exploring emergent quantum phenomena by chemical doping or synthesizing heterostructures.

\section{Associated Content}

\subsection{Supporting Information}

Crystallographic data from single-crystal XRD refinement; powder XRD data; additional STM/nc-AFM images; illustration of supercell; precession photos; neutron diffraction data (PDF).

\subsection{Notes}

The authors declare no competing financial interest.

\acknowledgement

This work was supported by the National Natural Science Foundation of China (No. 12474148, 12074426, 92477128 and 92580137), the National Key R\&D Program of China (MOST) (Grant No. 2023YFA1406500), the Fundamental Research Funds for the Central Universities and the Research Funds of Renmin University of China (No. 21XNLG27), This paper is an outcome of "Two-dimensional anisotropic series of materials FePd$_{2+x}$Te$_2$: a structural modulation study from the atomic scale to the mesoscopic scale " (RUC25QSDL128), funded by the Qiushi Academic Project of Renmin University of China in 2025.

	$^{\dag}$ These authors contributed equally to this work.

\begin{figure}
	\begin{center}
		TOC (For Table of Contents Only)
	\end{center}
	\centerline{\includegraphics[scale=0.5]{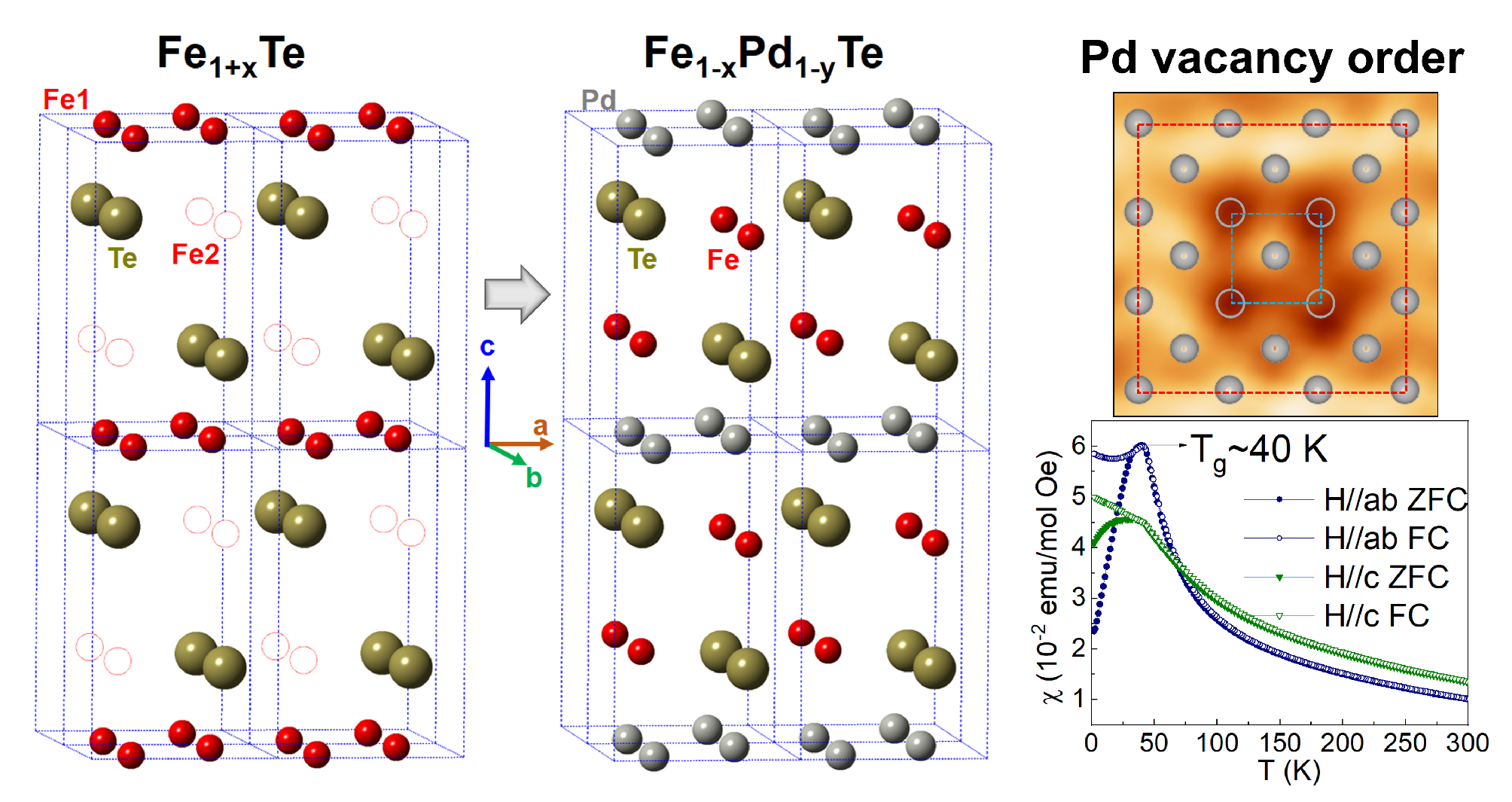}} \vspace*{-0.3cm}
\end{figure}

\end{document}